\documentclass[12pt]{article}

\usepackage{color}

\begin{document}
\title{CONFORMALLY FLAT SOURCES FOR THE LINET-TIAN SPACETIME}
\author{{Irene Brito$^1$\thanks{e-mail: ireneb@math.uminho.pt}, M. F. A. da Silva$^2$\thanks{e-mail: mfasnic@gmail.com}, Filipe C. Mena$^1$\thanks{e-mail: fmena@math.uminho.pt},
and N. O. Santos$^3$\thanks{e-mail: n.o.santos@qmul.ac.uk}}\\
{\small $^1$Centro de Matem\'atica, Universidade do Minho, 4710-057 Braga, Portugal.}\\
{\small $^2$Departamento de F\'{i}sica Te\'orica, Instituto de F\'isica,}\\
{\small Universidade do Estado do Rio de Janeiro, Rua S\~ao Francisco Xavier 524,}\\
{\small Maracan\~a, 20550-900, Rio de Janeiro, Brazil.}\\
{\small $^3$School of Mathematical Sciences, Queen Mary,}\\
{\small University of London, London E1 4NS, U.K.}\\
{\small $^3$Observatoire de Paris, Universit\'e Pierre et Marie Curie,} \\
{\small LERMA(ERGA) CNRS - UMR 8112, 94200 Ivry sur Seine, France.}}
\maketitle

\begin{abstract}
We investigate the matching, across cylindrical surfaces, of static cylindrically symmetric conformally flat spacetimes with a cosmological constant $\Lambda$, satisfying regularity conditions at the axis, to an exterior Linet-Tian spacetime. We prove that for $\Lambda\leq 0$ such matching is impossible. On the other hand, we show through simple examples that the matching is possible for $\Lambda>0$. We suggest a physical argument that might explain these results.
\end{abstract}

\section{Introduction}
The Levi-Civita spacetime \cite{LC} describes the vacuum field exterior to an infinite cylinder of matter. In its general form, it contains two independent parameters \cite{Bonnor,Bonnor1,Wang}, one, usually denoted by $\sigma$, describing the Newtonian energy per unit length, and another related to the angle defects. At first sight, global considerations in General Relativity seem to make cylindrical solutions to the Einstein field equations not so physically relevant: fields with cylindrical symmetry impose infinitely long sources, suggesting a peculiar physical situation. Nonetheless its importance cannot be underestimated, and under controlled circumstances, they provide a very good description of systems of physical interest (see e.g. \cite{Griffiths-Book}). Furthermore, Newtonian cylindrical models correspond well to observations \cite{Fujimoto,Hockney,Song}. In General Relativity, cylindrical solutions have been used to study various fields like cosmic strings \cite{Vilenkin,Silva}, exact models of rotation matched to different sources \cite{Mashhoon}, and models for extragalactic jets \cite{Opher,Herrerax,Gariel} and gravitational radiation \cite{Prisco}.
The generalization of the Levi-Civita spacetime to include a nonzero cosmological constant $\Lambda$ was obtained by Linet \cite{L} and Tian \cite{T} and it is shown by da Silva et al. \cite{Silva-etal} and Griffiths and Podolsky \cite{GP} that it changes the spacetime properties dramatically. The Linet-Tian (LT) solution has also been used to describe cosmic strings \cite{T,Bezerra,Bhattacharya} and, in \cite{ZofkaBicak}, static cylindrical shell sources have been found for the LT spacetime with negative cosmological constant. Considering this extensive interest in cylindrically symmetric solutions we assume worthwhile to analyze some further properties of LT spacetime.

In \cite{Nico}, while being studied conformally flat sources, it is proved a seemingly unexpected property that static cylindrical sources matched smoothly to the Levi-Civita spacetime exteriors do not admit conformally flat solutions. For spherical symmetry, there is the well known interior isotropic pressure and incompressible Schwarzschild solution, which is conformally flat \cite{Raychaudhuri}, matched to the Schwarzschild vacuum exterior spacetime. Senovilla and Vera \cite{Senovilla} obtained another disturbing result being the impossibility of the cylindrically symmetric Einstein-Straus model. In order to prove this impossibility they show that a Robertson-Walker spacetime, which is conformally flat, cannot be matched to any cylindrically symmetric static metric across a nonspacelike hypersurface preserving the symmetry. This result was subsequently generalised in \cite{Mars,Mena-Tavakol-Vera,Mars-Mena-Vera}. Another result that might be linked to this trend was obtained by Di Prisco et al. \cite{Prisco} and is the following: A cylindrically symmetric shear free collapsing anisotropic fluid can be matched to Einstein-Rosen spacetime as obtained in \cite{Prisco}, however, by considering that the exterior spacetime reduces to the static Levi-Civita spacetime, it imposes through its matching conditions, that the cylindrical source must be static\footnote{
We note that, in \cite{Bicak}, a static cylinder of an incompressible fluid has been matched to the Levi-Civita solution.}. We recall that a collapsing cylindrical shear free fluid, if it is isotropic, reduces to the conformally flat Robertson-Walker spacetime.

Here, we study static conformally flat solutions to an anisotropic fluid distribution bearing a non-zero cosmological constant and the possibility of matching them to the exterior LT spacetime.

The plan of the paper is as follows. In Section \ref{interior-ST}, we present the field equations for static anisotropic sources with non zero cosmological constant.
In Section 3, the matching conditions for the interior static anisotropic fluid to the LT exterior spacetime are given.
Section 4, is devoted to conformally flat solutions. The matching conditions when the interior spacetime is conformally flat and their consequences are analyzed in Section 5. We finish the paper with a conclusion suggesting a physical justification to our matching results.

We use latin indices $a,b,...=0,1,2,3$ and use units such that the speed of light $c=1$.

\section{Static cylindrically symmetric anisotropic sources with $\Lambda\ne 0$}
\label{interior-ST}

We consider a static cylindrically symmetric anisotropic fluid bounded by a cylindrical surface $S$ and with energy momentum tensor given by
\begin{equation}
T_{ab}=(\mu+P_r)V_{a}V_{b}+P_rg_{ab}+(P_z-P_r)S_{a}S_{b}+
(P_{\phi}-P_r)K_{a}K_{b}, \label{1a}
\end{equation}
where $\mu$ is the energy density, $P_r$, $P_z$ and $P_{\phi}$ are the principal stresses and $V_{a}$, $S_{a}$ and $K_{a}$ satisfy
\begin{equation}
V^{a}V_{a}=-1, \;\; S^{a}S_{a}=K^{a}K_{a}=1, \;\; V^{a}S_{a}=V^{a}K_{a}=S^{a}K_{a}=0. \label{1b}
\end{equation}
We assume for the interior to $S$ the general static cylindrically symmetric metric which can be written
\begin{equation}
ds^2=-A^2dt^2+B^2(dr^2+dz^2)+C^2d\phi^2, \label{1}
\end{equation}
where $A$, $B$ and $C$ are $C^2$-functions of $r$. To represent cylindrical symmetry, we impose the following ranges on the coordinates
\begin{equation}
-\infty< t<\infty, \;\; 0\leq r, \;\; -\infty <z<\infty, \;\; 0\leq\phi< 2\pi, \label{2}
\end{equation}
and $\phi=2\pi$ is identified with $\phi=0$.
We number the coordinates $x^0=t$, $x^1=r$, $x^2=z$ and $x^3=\phi$ and we choose the fluid being at rest in this coordinate system, hence from (\ref{1b}) and (\ref{1}) we have
\begin{equation}
V_{a}=-A\delta_{a}^0, \;\; S_{a}=B\delta_{a}^2, \;\; K_{a}=C\delta_{a}^3. \label{2a}
\end{equation}
For the Einstein field equations, $G_{ab}=\kappa T_{ab}-\Lambda g_{ab}$, where $\Lambda$ is the cosmological constant, with (\ref{1a}), (\ref{1}) and (\ref{2a}) we have the non zero components,
\begin{eqnarray}
G_{00}=-\left(\frac{A}{B}\right)^{2}\left[\left(\frac{B^{\prime}}{B}\right)'+
\frac{C^{\prime\prime}}{C}\right]=\kappa\bar{\mu}A^2, \label{2b}\\
G_{11}=\frac{A^{\prime}C'}{AC}+\left(\frac{A^{\prime}}{A}+\frac{C^{\prime}}{C}\right)
\frac{B^{\prime}}{B}=\kappa\bar{P}_{r} B^2, \label{2c}\\
G_{22}=\frac{A^{\prime\prime}}{A}+\frac{C^{\prime\prime}}{C}
+\frac{A^{\prime}}{A}\frac{C^{\prime}}{C}-\left(\frac{A'}{A}+\frac{C'}{C}\right)\frac{B'}{B}=\kappa\bar{P}_{z} B^2, \label{2d}\\
G_{33}=\left(\frac{C}{B}\right)^2 \left[\frac{A^{\prime\prime}}{A}+\left(\frac{B^{\prime}}{B}\right)'\right]=\kappa\bar{P}_{\phi} C^2, \label{2e}
\end{eqnarray}
where $\bar{\mu}=\mu+\Lambda/\kappa$, $\bar{P}_r=P_r-\Lambda/\kappa$, $\bar{P}_z=P_z-\Lambda/\kappa$,
$\bar{P}_{\phi}=P_{\phi}-\Lambda/\kappa$ and the primes stand for differentiation with respect to $r$.

The extension of the expression for the mass of an isolated system proposed by Tolman \cite{Tolman} and Whittaker \cite{Whittaker} to a non isolated system, bearing cylindrical symmetry, has been obtained by Israel \cite{Israel}. Other proposals for the mass per unit length exist, like by Marder \cite{Marder} and Vishveshwara and Winicour \cite{Vish}, but they proved to do not reproduce the expected Newtonian limit \cite{Wang}, while Israel's does. For this reason we use here the expression for mass per unit length obtained by Israel, which is
\begin{equation}
m=2\pi \int_{0}^{r_S} (\bar\mu +\bar{P}_r+\bar{P}_z+\bar{P}_\phi)\sqrt{-g}\; dr,\label{Wf}
\end{equation}
where $g$ is the determinant of the metric. Substituting (\ref{1}) and (\ref{2b})-(\ref{2e}) into (\ref{Wf}) one obtains
\begin{equation}
m=\frac{4\pi}{\kappa}\int_{0}^{r_S} (A'C)'dr,\label{Wm}
\end{equation}
and by considering the following regularity conditions on the axis \cite{Philbin}
\begin{equation}
\label{regularity}
A'(0)=B'(0)=C''(0)=C(0)=0, \;\; B(0)=C'(0)=1,
\end{equation}
equation (\ref{Wm}), at $r=r_S$, becomes
\begin{equation}
m\stackrel{S}{=}\frac{4\pi}{\kappa}A^{\prime}C, \label{Wm1}
\end{equation}
where $\stackrel{S}{=}$ denotes equality on $S$.

Since we are concerned with conformally flat sources for the LT spacetime, we need in the sequel the square of the magnitude of the Weyl tensor ${\mathcal C}^2=C^{abcd}C_{abcd}$, which can be written with the aid of the field equations (\ref{2b})-(\ref{2e}) as
\begin{eqnarray}
{\mathcal C}^2=\frac{2}{3}\left\{\left[\kappa(\bar\mu+\bar{P}_z)+\frac{2}{B^2}(\beta-\gamma)\right]^2
+\left[\kappa(\bar\mu+\bar{P}_{\phi})
+\frac{2}{B^2}(\beta-\alpha)\right]^2\right. \nonumber\\
\left.+\left[\kappa(\bar{P}_z-\bar{P}_{\phi})+\frac{2}{B^2}(\alpha-\gamma)\right]^2\right\}, \label{13a}
\end{eqnarray}
where
\begin{equation}
\frac{A^{\prime}}{A}\frac{B^{\prime}}{B}=\alpha, \;\;
\frac{B^{\prime}}{B}\frac{C^{\prime}}{C}=\beta, \;\;
\frac{A^{\prime}}{A}\frac{C^{\prime}}{C}=\gamma. \label{13}
\end{equation}

\section{LT spacetime and matching conditions}
\label{Linet-Tian}
In this section, we match the interior spacetime, bounded by the surface $S$ and given by the metric (\ref{1}), to an exterior described by the LT spacetime containing the cosmological constant. The generalized static cylindrically symmetric Levi-Civita metric with non zero $\Lambda$, given
in its usual form by the LT metric \cite{L,T} is
\begin{eqnarray}
\label{exterior-metric}
ds^{2+}=-a^2Q^{2/3}P^{-2(1-8\sigma+4\sigma^2)/3\Sigma}dt^2+d\rho^2
+b^2Q^{2/3}P^{-2(1+4\sigma-8\sigma^2)/3\Sigma}dz^2 \nonumber\\
+c^2Q^{2/3}P^{4(1-2\sigma-2\sigma^2)/3\Sigma}d\phi^2, \label{16}
\end{eqnarray}
where $\Sigma=1-2\sigma+4\sigma^2$, and for $\Lambda<0$,
\begin{equation}
Q(\rho)=\frac{1}{\sqrt{3|\Lambda|}}\sinh(2R), \;\;
P(\rho)=\frac{2}{\sqrt{3|\Lambda|}}\tanh R, \label{17}
\end{equation}
with
\begin{equation}
R=\frac{\sqrt{3|\Lambda|}}{2}\rho, \label{17a}
\end{equation}
and $a$, $b$, $c$ and $\sigma\ge 0$ are real constants.
The case $\Lambda>0$ is obtained by replacing the hyperbolic functions by trigonometric ones \cite{L,T}.
% and, for that case, the calculations in this section remain also valid throughout this section.
The coordinates $t$, $z$ and $\phi$ in (\ref{exterior-metric}) can be taken the same as in (\ref{1}) and with the same ranges (\ref{2}). The radial coordinates $r$ and $\rho$ are not necessarily continuous on $S$ as we see below by applying the junction conditions. The constants $a$ and $b$ can be removed by scale transformations (although we don't do this ahead in order to use these constants as free parameters for the matching), while $c$ cannot be transformed away if we want to preserve the range of $\phi$. The constant $\sigma$ represents the Newtonian mass per unit length.

Following Darmois junction conditions \cite{Darmois} we impose that, on the surface $S$, the first and second fundamental forms which $S$ inherits from the interior metric (\ref{1}) and from the exterior metric (\ref{16}) are equal, hence we obtain the following two sets of equations on $S$,
\begin{eqnarray}
\label{match1}
A\stackrel{S}{=}a\,Q^{1/3}P^{-(1-8\sigma+4\sigma^2)/3\Sigma}, \label{15a}\\
B\stackrel{S}{=}b\,Q^{1/3}P^{-(1+4\sigma-8\sigma^2)/3\Sigma}, \label{16a}\\
C\stackrel{S}{=}c\,Q^{1/3}P^{2(1-2\sigma-2\sigma^2)/3\Sigma}, \label{17a}
\end{eqnarray}
and\footnote{In the case $\Lambda>0$, the hyperbolic functions in (\ref{21})-(\ref{23a}) are substituted by trigonometric ones.}
\begin{eqnarray}
\label{match3}
\frac{A^{\prime}}{AB}\stackrel{S}{=}\sqrt{\frac{|\Lambda|}{3}}\,\frac{\Sigma(\cosh^2R-1)+3\sigma}
{\Sigma\sinh R\cosh R}, \label{21}\\
\frac{B^{\prime}}{B^2}\stackrel{S}{=}\sqrt{\frac{|\Lambda|}{3}}\,
\frac{\Sigma(\cosh^2R-1)-3\sigma(1-2\sigma)}{\Sigma\sinh R\cosh R}, \label{22}\\
\frac{C^{\prime}}{BC}\stackrel{S}{=}\sqrt{\frac{|\Lambda|}{3}}\,\frac{2\Sigma(\cosh^2R-1)+
3(1-2\sigma)}{2\Sigma\sinh R\cosh R}.\label{23a}
\end{eqnarray}
By replacing the matching conditions (\ref{15a})-(\ref{23a}) in (\ref{2c}) we get, for both cases $\Lambda<0$ and $\Lambda>0$, that
\begin{equation}
\label{pressure}
\bar{P}_r\stackrel{S}{=}-\Lambda \;\; \mbox{or} \;\; P_r\stackrel{S}{=}0,
\end{equation}
as expected.
The mass per unit length (\ref{Wm1}) with (\ref{15a})-(\ref{21}) and considering the gravitational coupling constant $G=1$, then $\kappa=8\pi$, can be written as
\begin{equation}
m\stackrel{S}{=}m_{LC}+\frac{abc}{3}\sinh^2R, \label{MM1}
\end{equation}
where the mass per unit length for the Levi-Civita metric, with $\Lambda=0$, is
\begin{equation}
m_{LC}=abc\frac{\sigma}{\Sigma}, \label{MM1a}
\end{equation}
thus showing that the presence of $\Lambda<0$ increases the mass per unit length. However, for $\Lambda>0$ we obtain
\begin{equation}
m\stackrel{S}{=}m_{LC}-\frac{abc}{3}\sin^2R, \label{MM2}
\end{equation}
producing an opposite effect, diminishing the mass per unit length. In the Conclusion we consider these results as a possible justification for the possibility, or impossibility, of matching conformally flat interior spacetimes to LT exteriors.

\section{Conformally flat interior sources}
The conformally flat spacetime solution, where all Weyl tensor components vanish, $C_{abcd}=0$, for (\ref{1}) with the regularity conditions
(\ref{regularity}) satisfied produces \cite{Nico}
\begin{eqnarray}
A=a_1\cosh(a_2r)B, \label{I5} \\
C=\frac{1}{a_2}\sinh(a_2 r)B, \label{I4}
\end{eqnarray}
where $a_1\ne 0$ and $a_2\ne 0$ are integration constants, and by rescaling $t$ we can assume $a_1=1$.

The interpretation of $a_2$ can be given in the following way.
From (\ref{I5}) and (\ref{I4}) we can write
\begin{eqnarray}
\frac{A^{\prime}}{A}=\frac{B^{\prime}}{B}+a_2\tanh(a_2r), \label{II6a}\\
\frac{C^{\prime}}{C}=\frac{B^{\prime}}{B}+a_2\coth(a_2r), \label{II6}
\end{eqnarray}
and with (\ref{13a}) and (\ref{13}) it follows,
\begin{eqnarray}
\kappa({\bar\mu}+{\bar P}_z)=\frac{2a_2}{B^2}\left[\tanh(a_2r)\frac{B^{\prime}}{B}+a_2\right], \label{II7}\\
\kappa({\bar\mu}+{\bar P}_{\phi})=\frac{2a_2}{B^2}\left[\tanh(a_2r)-\coth(a_2r)\right]\frac{B^{\prime}}{B}, \label{II8}
\end{eqnarray}
producing
\begin{equation}
\tanh^2(a_2r)=\frac{2a_2^2-\kappa({\bar\mu}+{\bar P}_z)B^2}{2a_2^2+\kappa({\bar P}_{\phi}-{\bar P}_z)B^2}. \label{II9}
\end{equation}
At the centre of the source, $r=0$, considering the regularity conditions (\ref{regularity}) we have from (\ref{II9})
\begin{equation}
2a_2^2=\kappa({\bar\mu}_0+{\bar P}_{z0}). \label{II10}
\end{equation}

\section{Interior static conformally spacetime matched to exterior LT spacetime}

We start by considering the matching on $S$ for $\Lambda<0$. Then, (\ref{II6}) with (\ref{22}) and (\ref{23a}) becomes
\begin{equation}
\frac{a_2}{B}\coth(a_2r)\stackrel{S}{=}\frac{\sqrt{3|\Lambda|}}{2\Sigma}\frac{1-4\sigma^2}{\sinh R\cosh R}. \label{I8}
\end{equation}
From the equality of the interior and exterior first fundamental forms on $S$ we have $B^2dr^2\stackrel{S}{=}d\rho^2$ which, using (\ref{16a}), leads to the relation
\begin{equation}
r\stackrel{S}{=}\frac{1}{b}\int_0^{\rho_S}\frac{d\rho}{Q^{1/3}P^{-(1+4\sigma-8\sigma^2)/3\Sigma}}, \label{I10}
\end{equation}
with $b>0$.
Then, using the equality
\begin{eqnarray}
\frac{2\Sigma}{\sqrt{3|\Lambda|}b}\frac{\sinh R\cosh R}{Q^{1/3}P^{-(1+4\sigma-8\sigma^2)/3\Sigma}}=
\frac{4\Sigma}{3b}\int\frac{\sinh^2R\,d\rho}{Q^{1/3}P^{-(1+4\sigma-8\sigma^2)/3\Sigma}} \nonumber\\
+\frac{1}{b}\int\frac{d\rho}{Q^{1/3}P^{-(1+4\sigma-8\sigma^2)/3\Sigma}}, \label{I9}
\end{eqnarray}
at the boundary S, (\ref{I8}) becomes
\begin{equation}
a_2\left(r+\frac{4\Sigma}{3b}\int_0^{\rho_S}\frac{\sinh^2R\,d\rho}{Q^{1/3}P^{-(1+4\sigma-8\sigma^2)/3\Sigma}}\right)
\coth(a_2r)\stackrel{S}{=}1-4\sigma^2. \label{I11}
\end{equation}
Since the left hand side of (\ref{I11}) is always bigger than $1$ this condition can never be satisfied.

When $\Lambda=0$, equation (\ref{I11}) reduces to
\begin{equation}
a_2r\coth(a_2r)\stackrel{S}{=}1-4\sigma^2, \label{I12}
\end{equation}
obtained in \cite{Nico} for the case of a Levi-Civita exterior, which again shows the impossibility of matching a cylindrical conformally flat interior spacetime to a  Levi-Civita exterior. Then we can state the following:

{\it It is impossible to match any conformally flat static cylindrically symmetric interior spacetime (\ref{I5}) and (\ref{I4}) satisfying the regularity conditions (\ref{regularity}) to an exterior LT spacetime, with $\Lambda<0$, or to an exterior Levi-Civita spacetime, with $\Lambda=0$, across a timelike cylindrical hypersurface $S$.}

 For $\Lambda>0$,
the corresponding equation to (\ref{I11}) becomes
\begin{equation}
a_2\left(r-\frac{4\Sigma}{3b}\int_0^{\rho_S}\frac{\sin^2R\,d\rho}{Q^{1/3}P^{-(1+4\sigma-8\sigma^2)/3\Sigma}}\right)
\coth(a_2r)
\stackrel{S}{=}1-4\sigma^2, \label{J12}
\end{equation}
and since the left hand side is less than 1 it does not discard, a priori, conformally flat sources matched to the LT spacetime with $\Lambda>0$.

Now, we give simple examples\footnote{We note that in \cite{GP}, the LT spacetime with $\Lambda>0$ was matched to the toroidal Einstein static universe.} for which a conformally flat interior source can be matched to an exterior LT spacetime with $\Lambda>0$.

\subsection{$\bar P_r=\bar P_z$ or $\bar P_z=\bar P_{\phi}$}

 In this case, the solution of (\ref{2b})-(\ref{2e}) with (\ref{I5}) and (\ref{I4}) can be easily demonstrated to be
\begin{equation}
B=\frac{1}{\cosh(a_2r)}, \label{J13}
\end{equation}
and
\begin{equation}
\bar{P}_r=\bar{P}_z=\bar{P}_{\phi}=-\frac{\bar\mu}{3}=-\frac{a_2^2}{\kappa}. \label{J14}
\end{equation}
By matching this solution on $S$ to the exterior LT spacetime we have from (\ref{pressure}),
\begin{equation}
P_r=P_z=P_{\phi}=0, \;\; \mu=2\frac{\Lambda}{\kappa}, \label{J15}
\end{equation}
reducing the interior solution to the Einstein static universe.

The junction conditions (\ref{15a})-(\ref{23a}) for $\Lambda>0$ and (\ref{J13}) become
\begin{eqnarray}
1\stackrel{S}{=}a\,Q^{1/3}P^{-(1-8\sigma+4\sigma^2)/3\Sigma}, \label{M1}\\
\frac{1}{\cosh(a_2r)}\stackrel{S}{=}b\,Q^{1/3}P^{-(1+4\sigma-8\sigma^2)/3\Sigma}, \label{M2}\\
\frac{1}{a_2}\tanh(a_2r)\stackrel{S}{=}c\,Q^{1/3}P^{2(1-2\sigma-2\sigma^2)/3\Sigma}, \label{M3}
\end{eqnarray}
and
\begin{eqnarray}
0\stackrel{S}{=}\sqrt{\frac{\Lambda}{3}}\frac{3\sigma-\Sigma\sin^2R}{\Sigma\sin R\cos R}, \label{M4}\\
a_2\sinh(a_2r)\stackrel{S}{=}\sqrt{\frac{\Lambda}{3}}\frac{3\sigma(1-2\sigma)+\Sigma\sin^2R}{\Sigma\sin R\cos R}, \label{M5}\\
\frac{a_2}{\sinh(a_2r)}\stackrel{S}{=}\sqrt{\frac{\Lambda}{3}}\frac{3(1-2\sigma)-2\Sigma\sin^2R}{2\Sigma\sin R\cos R}.
\label{M6}
\end{eqnarray}
From (\ref{M4})-(\ref{M6}) we have $a_2^2=\Lambda$, as can be obtained too from (\ref{II10}), and
\begin{equation}
\sin^2R\stackrel{S}{=}\frac{3\sigma}{\Sigma}, \;\; \sinh^2(\sqrt{\Lambda}r)\stackrel{S}{=}\frac{4\sigma(1-\sigma)}{1-4\sigma}, \label{M7}
\end{equation}
where $0\leq\sin R_S\leq 1$ and
$0\leq\sinh(\sqrt{\Lambda}r_S)<\infty$ are satisfied by $0<\sigma<
1/4$. While (\ref{M1})-(\ref{M3}) with (\ref{M7}) define the
exterior parameters $a$, $b$ and $c$ in terms of $\Lambda$ and
$\sigma$.

Hence, it is possible to match a conformally flat static cylindrically symmetric interior spacetime (\ref{I5}) and (\ref{I4}), satisfying regularity conditions (\ref{regularity}), to an exterior LT spacetime with $\Lambda>0$, across a timelike cylindrical hypersurface $S$.

We call attention to the fact that the LT spacetime with $\Lambda>0$ has, besides the singularity at $\rho=0$ where we placed the source, another singularity at $\rho=\pi/\sqrt{3\Lambda}$ where another source has to be placed. In that case, the matching is still possible by substituting the cylindrical region by a toroidal one following the methods of \cite{GP}.

\subsection{$\bar P_r=\bar P_{\phi}$}
In this case, the solution of (\ref{2b})-(\ref{2e}) with
(\ref{I5}) and (\ref{I4}) is
\begin{eqnarray}
\label{exact2} B=\frac{1}{a_4[\cosh(a_2r)-1]+1}
\end{eqnarray}
where $a_4 \neq 0$ is a constant. We note that if $a_4=1$, the function $B$ corresponds to the solution (\ref{J13}). The density and pressures have the following form
\begin{eqnarray}
&&\bar{\mu}=2\;a_2^{2}\;a_4\;[(1-a_4)\cosh(a_2 r)+a_4 +1]-a_2^{2}\\
&& \bar{P_r}=2a_2^{2}a_4[(a_4 -1)\tanh(a_2 r)\sinh(a_2 r)-1]+a_2^{2}\\
&& \bar{P_z}=2a_2^{2}a_4\left[\frac{1-a_4}{\cosh(a_2 r)}+a_4 -3\right]+3a_2^{2}
\end{eqnarray}
In this case, the matching conditions (\ref{15a})-(\ref{23a}) read
\begin{eqnarray}
\frac{\cosh(a_2 r)}{a_4[\cosh(a_2 r)-1]+1}\stackrel{S}{=}a\,Q^{1/3}P^{-(1-8\sigma+4\sigma^2)/3\Sigma}, \label{Mc31}\\
\frac{1}{a_4[\cosh(a_2 r)-1]+1}\stackrel{S}{=}b\,Q^{1/3}P^{-(1+4\sigma-8\sigma^2)/3\Sigma}, \label{Mc32}\\
\frac{\sinh(a_2 r)}{ a_2 [a_4[\cosh(a_2 r)-1]+1]}\stackrel{S}{=}c\,Q^{1/3}P^{2(1-2\sigma-2\sigma^2)/3\Sigma}, \label{Mc33}
\end{eqnarray}
and
%\begin{eqnarray}
%-3a_2(a_4-1)\tanh(a_2 r)\stackrel{S}{=}\frac{Q'}{Q}-\left(\frac{1-8\sigma+4\sigma^2}{1-2\sigma+4\sigma^2}\right)\frac{P'}{P}, \label{Mc44}\\
%-3a_2 a_4\sinh(a_2 r)\stackrel{S}{=}\frac{Q'}{Q}-\left(\frac{1+4\sigma-8\sigma^2}{1-2\sigma+4\sigma^2}\right)\frac{P'}{P}, \label{Mc45}\\
%-3\frac{a_2[\cosh(a_2 r)(a_4 -1)-a_4]}{\sinh(a_2 r)}\stackrel{S}{=}\frac{Q'}{Q}+2\left(\frac{1-2\sigma-2\sigma^2}{1-2\sigma+4\sigma^2}\right)\frac{P'}{P}.
%\label{Mc46}
%\end{eqnarray}
%
\begin{eqnarray}
a_2(1-a_4)\tanh(a_2
r)\stackrel{S}{=}\sqrt{\frac{\Lambda}{3}}\,\frac{3\sigma-\Sigma
\sin^2R }
{\Sigma\sin R\cos R}, \label{Mc44n}\\
a_2 a_4\sinh(a_2 r)\stackrel{S}{=}\sqrt{\frac{\Lambda}{3}}\,
\frac{3\sigma(1-2\sigma)+\Sigma\sin^2R}{\Sigma\sin R\cos R}, \label{Mc45n}\\
\frac{a_2[\cosh(a_2 r)(1-a_4)+a_4]}{\sinh(a_2
r)}\stackrel{S}{=}\sqrt{\frac{\Lambda}{3}}\,\frac{3(1-2\sigma)-
2\Sigma\sin^2R }{2\Sigma\sin R\cos R}. \label{Mc46n}
\end{eqnarray}
From (\ref{Mc44n})-(\ref{Mc46n}) we obtain
\begin{eqnarray}
\label{sig1} \sin^2 R \stackrel{S}{=} \frac{3
\sigma}{\Sigma}\left[1+ \frac{2(a_4 -1) (1-\sigma)
\sqrt{1-4\sigma}}{a_4 \sqrt{1-4\sigma^2}- (a_4
-1)\sqrt{1-4\sigma}}\right]
\end{eqnarray}
and
\begin{eqnarray}
\label{sig6} \sinh^2(a_2
r)\stackrel{S}{=}\frac{4\sigma(1-\sigma)}{1-4\sigma},
\end{eqnarray}
as well as (which also follows from (\ref{pressure}))
\begin{eqnarray}
\label{press} a_2^2 \stackrel{S}{=}\Lambda \left[2a_4-1-2a_4(a_4
-1)\frac{4\sigma(1-\sigma)}{\sqrt{(1-4\sigma)(1-4\sigma^2)}}\right]^{-1}.
\end{eqnarray}
%Substituting $Q'/Q$ obtained from (\ref{Mc45}) in (\ref{Mc44}) and (\ref{Mc46}) and then, subtracting the second from the first equation one gets
%\begin{eqnarray}
%\label{PPP}
%\frac{P'}{P}\stackrel{S}{=}a_2\frac{a_4 (1-\cosh(a_2 r))-1}{\cosh(a_2 r)\sinh(a_2 r)}\left(\frac{1-2\sigma+4\sigma^2}{4\sigma-1}\right).
%\end{eqnarray}
%Then, it follows from (\ref{Mc44}) or (\ref{Mc46}) that
The inequality $0\le \sin^2 R_S\le 1$ in (\ref{sig1}) and the
positivity of the right hand side of (\ref{press}), for any $0<\sigma<1/4$, are
satisfied if $1/2\le a_4\le 1$.

We conclude that, in this case, the matching is possible in the
following sense:

For any $1/2\le a_4\le 1$, $0<\sigma<1/4$ and $\Lambda>0$, the
parameter $a_2$ is fixed by (\ref{press}) while  $\rho_S$ and $r_S$
are determined from (\ref{sig1}) and (\ref{sig6}). In turn,
(\ref{Mc31})-(\ref{Mc33}) fix the exterior parameters $a, b$ and
$c$. If $a_4=1$, this solution reduces to the example of the
previous section.

\section{Conclusion}
The main result obtained here is that it is not possible to match a static interior cylindrically symmetric  conformally flat spacetime smoothly across a cylindrical surface to an exterior given by the LT spacetime, when $\Lambda<0$, or, by the Levi-Civita spacetime when $\Lambda=0$. For $\Lambda>0$, it is possible to perform such matching as we showed with  two examples.

We also showed, that the mass per unit length is increased by the presence of $\Lambda<0$, while it is diminished by $\Lambda>0$.

The Levi-Civita spacetime does not possess any horizons, which may seem to indicate, according to our understanding of black hole formation, that there is an upper limit allowed by the mass per unit length. This limit is always below the critical linear mass above which horizons may be formed \cite{Wang,Lathrop}. The fact that conformally flat spacetimes cannot be matched to Levi-Civita might be physically justified from the fact that sources producing these spacetimes have linear masses higher than this limit. If this is the case, then the inclusion of $\Lambda<0$ would further unbalance this limit since, from (\ref{MM1}), the mass per umit length would be further increased.
On the other hand, for $\Lambda>0$, we see that the matching is possible and, from (\ref{MM2}), the corresponding mass per unit length is diminished as compared to the Levi-Civita linear mass. This fact might suggest that, in this case, the linear mass is sufficiently diminished as compared to the critical mass limit.

\section*{Acknowledgments}
We thank the referees for useful criticisms. IB and FM thank CMAT,
Univ. Minho, for support through the FEDER Funds - "Programa
Operacional Factores de Competitividade  COMPETE" and FCT Project
Est-C/MAT/UI0013/2011. FM is supported by FCT projects
PTDC/MAT/108921/2008 and CERN/FP/116377/2010. MFAdaSilva
acknowledges the financial support from FAPERJ (no.
E-26/171.754/2000, E-26/171.533.2002, E-26/170.951/2006,
E-26/110.432/2009 and E-26/111.714/2010), Conselho Nacional de
Desenvolvimento Cient\'{i}fico e Tecnol\'ogico - CNPq - Brazil (no.
450572/2009-9, 301973/2009-1 and 477268/2010-2) and Financiadora de
Estudos e Projetos - FINEP - Brazil.

\end{document}